\documentclass[aps,prd,preprintnumbers,nofootinbib]{revtex4}
\usepackage[utf8]{inputenc}
\usepackage{bm}
\usepackage[normalem]{ulem}
\usepackage{latexsym}
\usepackage{color}
\usepackage{dcolumn}
\usepackage{amsmath,amsfonts,amssymb}
\usepackage{graphicx,epsfig}
\usepackage{psfrag}
\usepackage{amsthm}
\usepackage[pdftex]{hyperref}
\usepackage[colorinlistoftodos]{todonotes}
\usepackage{amsmath}
\usepackage[toc,title,page]{appendix}
\usepackage{natbib}
\def\be{\begin{eqnarray}}
\def\ee{\end{eqnarray}}

\begin{document}

\title{Cosmological Complexity}
\author{Arpan Bhattacharyya} \email{abhattacharyya@iitgn.ac.in}
\affiliation{
Indian Institute of Technology,
Gandhinagar,Gujarat 382355, India}
\author{Saurya Das}\email{saurya.das@uleth.ca}
\affiliation{Theoretical Physics Group, \\
Department of Physics and Astronomy,\\
University of Lethbridge, 4401 University Drive,\\
Lethbridge, Alberta T1K 3M4, Canada}
\author{S. Shajidul Haque}\email{shajid.haque@uct.ac.za}
\affiliation{High Energy Physics, Cosmology \& Astrophysics Theory Group \\and The Laboratory for Quantum Gravity \& Strings,\\
Department of Mathematics and Applied Mathematics,\\
University of Cape Town, South Africa} 
\author{Bret Underwood}\email{bret.underwood@plu.edu}
\affiliation{Department of Physics,\\
Pacific Lutheran University,\\
Tacoma, WA 98447}

\date{\today}

\vspace{-2cm}
\begin{abstract}
We compute the quantum circuit complexity of the evolution of scalar curvature perturbations on expanding backgrounds, using the language of squeezed vacuum states. In particular, we construct a simple cosmological model consisting of an early-time period of de Sitter expansion followed by a radiation-dominated era and track the evolution of complexity throughout this history. During early-time de Sitter expansion the complexity grows linearly with the number of e-folds for modes outside the horizon. The evolution of complexity also suggests that the Universe behaves like a chaotic system during this era, for which we propose a scrambling time and Lyapunov exponent. During the radiation-dominated era, however, the complexity decreases until it ``freezes in'' after horizon re-entry, leading to a ``de-complexification'' of the Universe.

\end{abstract}

\maketitle
\section{Introduction}
In recent years, Quantum Information Theory has played the role of a melting pot for various branches of physics. In the context of high energy theory, a motivation to understand the application of complexity to quantum field theory arises from attempts to apply the AdS/CFT duality in certain black hole settings. In particular, it is notoriously difficult to probe physics behind the horizon of a black hole. It has been observed that although the entanglement entropy of an eternal AdS black hole saturates as it thermalizes \cite{Hartman}, the size of the Einstein-Rosen bridge continues to increase with time. 
Motivated by this observation,
Susskind et.~al.~\cite{Susskind,Susskind1,Susskind2,Susskind3,Susskind4,Susskind5,Susskind6} have proposed new probes on the gravity side for the inner region beyond the black hole horizon. 
One probe is given by the volume of a maximal codimension-one bulk surface extending to the boundary of AdS spacetime \cite{Susskind,Susskind1,Susskind2,Susskind3,Susskind4}. There is a second proposal, where the probe is the action defined on the Wheeler-DeWitt (WDW) patch \cite{Susskind5,Susskind6}. 
Both of these quantities have the potential to probe physics behind the horizon. It is conjectured that these two objects are dual to the so-called ``complexity" of the dual field theory state. For this reason these proposals are known as the CV (complexity = volume) \cite{Susskind3} and CA (complexity = action) \cite {Susskind5} conjectures, respectively. These conjectures have opened up a completely new line of research that relates high energy theory and condensed matter physics with quantum information theory at the centre, e.g \cite{Couch,Swingle:2017zcd,Bhattacharyya:2018wym,Ham, Bolognesi:2018ion}\footnote{This list is by no means complete. Interested readers are requested to check the citations and references of these papers.}.

The holographic proposals mentioned above connect a probe on the gravity side with a concept in quantum information theory called quantum complexity \cite{watrous}. More specifically we will be focusing on circuit complexity. Circuit complexity is the minimum number of unitary operators (also known as quantum gates) that are required to construct the desired target state from a suitable reference state. 
For Gaussian states, this can either be computed by working directly with the wavefunctions in the position basis \cite{MyersCC,MyersCCa,MyersCC1} or using a covariance matrix \cite{MyersCC2,MyersCC3,MyersCC4,MyersCC5,MyersCC6,MyersCC7}. 
In both these cases, the quantum complexity is typically computed using a geometric technique pioneered by Nielsen \cite{NL1,NL2,NL3}. Alternatively, it has also been proposed that the quantum complexity might be computed using Fubini-Study distance \cite{MyersCC8}.
It has been shown in \cite{AB} (especially in the context of certain time evolution) that out of all these methods, the quantum complexity computed using wavefunction might be the most sensitive one to the underlying physics.

Over the past few years, circuit complexity has enjoyed a wide range of applications. 
For instance, quantum complexity may be a possible diagnostic for quantum chaos, and is now considered as an integral part of the web of diagnostics for quantum chaos \cite{qchaospre, qchaosprea,qchaos,qchaos1,qchaos2,qchaos3,qchaos4}. It was highlighted in \cite{qchaos1} that circuit complexity can provide essential information (such as the scrambling time, Lyapunov exponent, etc.) about a quantum chaotic system. 
In \cite{qchaos1}, an inverted harmonic oscillator model was used to establish the chaotic features of complexity and compared them with the information one can obtain from the out-of-time-order correlators. The time scale when the complexity starts to grow was identified as the scrambling time and the slope of the linear portion behaves as the Lyapunov exponent.

In this paper we use this in the field of cosmology. More explicitly, we apply the notion of circuit complexity to scalar cosmological perturbations on an expanding 
Friedmann-Lemaitre-Robertson-Walker (FLRW)
background. Scalar perturbations on an expanding background can naturally be described with the formalism of squeezed quantum states: when a mode exits the horizon it becomes highly squeezed, while a mode inside the horizon has its squeezing ``frozen in''\cite{Grishchuk,Albrecht}.
We will choose the ground state while the mode is inside the horizon as our reference state, and study complexity for a target state consisting of the time-evolved cosmological perturbation on the expanding background. For simplicity we consider a simple model 
consisting of a period of de Sitter (dS) expansion followed by radiation-dominated expansion, as a proxy for inflation followed by reheating.

This approach gives us interesting behaviors for the complexity of cosmology at different epochs. We find that during dS expansion, the complexity is proportional to the number of e-folds for a super-horizon mode. The exponential growth as in \cite{qchaos1}, suggests that during the de Sitter regime the complexity grows as in an unstable (chaotic) system. Moreover, one can also identify the scrambling time scale for this chaotic regime and the Lyapunov exponent. During the subsequent radiation phase the Universe de-complexifies, even though the squeezing of the perturbation continues, and eventually the complexity ``freezes in'' once the mode re-enters the horizon.

The organization of the paper is as follows. In Section \ref{sec:Inverted} we will use the inverted harmonic oscillator model to get insights about our approach and to establish our tools and techniques. In Section \ref{sec:SqueezedCosmo} we review the cosmological scalar perturbations and the origin of the squeezed states and the various solutions. In Section \ref{sec:Complexity} we discuss the complexity for this squeezed states and discuss the evolution of complexity and its implications. We conclude with a discussion and future directions.

\section{Inverted Harmonic Oscillator}
\label{sec:Inverted}

To begin, we will introduce the main techniques and concepts used throughout the paper through the example of the inverted harmonic oscillator. 
Since a super-horizon scalar cosmological perturbation behaves like an inverted harmonic oscillator at large scales, the intuition we develop here will be useful for our later analysis.

The inverted harmonic oscillator is defined by a Hamiltonian with a ``wrong sign'' of the restoring force (with unit mass) \cite{Barton}:
\be
\hat H = \frac{1}{2} \hat p^2 - \frac{1}{2} k^2 \hat x^2.
\label{InvertHpq}
\ee

Using the raising and lowering operators based on the non-inverted harmonic oscillator
\be
\hat x = \frac{1}{\sqrt{2k}} \left(\hat a^\dagger + \hat a\right), \hspace{.4in} \hat p = i \sqrt{\frac{k}{2}} \left(\hat a^\dagger - \hat a\right)\, ,
\ee
the inverted Hamiltonian (\ref{InvertHpq}) becomes
\be
\hat H = - \frac{k}{2} \left(\hat a^2 + \hat a^\dagger{}^2\right)\, .
\label{InvertH}
\ee

If the system starts in the ``vacuum state'' annihilated by the lowering operator
\be
\hat a |0\rangle = 0\, ,
\ee
then it will naturally evolve into a squeezed state at later times.
In particular, the unitary evolution $\hat {\mathcal U}$ of a state can be parameterized as \cite{Grishchuk,Albrecht}
\be
\hat {\mathcal U} = \hat {\mathcal S}(r,\phi) \hat {\mathcal R}(\theta)\, .
\ee
where $\hat {\mathcal R}$ is the ``rotation operator,'' defined as
\be
\hat {\mathcal R}(\theta) \equiv {\rm exp}\left[-i\theta(t) (\hat a^\dagger \hat a + \hat a \hat a^\dagger)\right]
\ee
in terms of the rotation parameter $\theta(t)$,
and $\hat {\mathcal S}(r,\phi)$ is the ``squeezing operator,'' defined as
\be
\hat {\mathcal S}(r,\phi) \equiv {\rm exp} \left[\frac{r(t)}{2}\left(e^{-2i\phi}\hat a^2 - e^{2i\phi} \hat a^\dagger{}^2\right)\right]
\ee
in terms of the squeezing parameter $r(t)$ and squeezing angle $\phi(t)$.
In what follows, the rotation operator and rotation parameter will not play an important role, so we will drop them from our subsequent analysis.

The action of the rotation operator produces an irrelevant phase; however, the action of the squeezing operator results in a single mode squeezed vacuum state \cite{book}:
\be
|\Psi(t)\rangle = \hat {\mathcal S}(r,\phi) |0\rangle = \frac{1}{\sqrt{\cosh r}}\sum_{n=0}^\infty (-1)^n\ e^{-2in\phi} \tanh^n r\  \frac{\sqrt{(2n)!}}{2^n n!}\ |2n\rangle\, .
\label{InvertedPsi}
\ee
To understand the importance of the squeezing angle and squeezing parameter, consider the combinations 

\be
\hat q_+ &\equiv & \hat p \sin \phi + k\ \hat x \cos \phi\, ; \\
\hat q_- &\equiv & \hat p \cos\phi - k\ \hat x \sin \phi\, .
\ee
The uncertainty for these new variables is \cite{Albrecht}
\be
\label{InvertedSqueeze1}
\Delta q_+^2 &=& \langle \Psi(t)|\hat q_+^2 |\Psi(t) \rangle = \frac{1}{2} e^{-2r}\, ; \\
\Delta q_-^2 &=& \langle \Psi(t)|\hat q_-^2 |\Psi(t) \rangle = \frac{1}{2} e^{2r}\, .
\label{InvertedSqueeze2}
\ee
This clearly shows the origin of the term ``squeezed states'': the wavefunction $|\Psi(t)\rangle$ is squeezed with a small uncertainty in the $\hat q_+$ direction, with a correspondingly large uncertainty in the $\hat q_-$ direction, so that the uncertainty relation is still saturated $\Delta q_+ \Delta q_- = 1/2$.
The squeezing angle $\phi$ determines the angle in phase space at which
the squeezing occurs.

It is straightforward to insert (\ref{InvertedPsi}) into the Schr\"odinger equation
\be
i \frac{d}{dt} |\Psi(t)\rangle = \hat H |\Psi(t) \rangle
\ee
to obtain the squeezing equations of motion
\be
\dot r &=& k \sin( 2\phi)\, ; \nonumber \\
\dot \phi &=& k \coth(2r) \cos(2\phi)\, . \label{InvertedEOM}
\ee
It is easy to see that these equations have a solution in which the squeezing grows with time along a constant squeeze angle
\be
r(t) = k t, \hspace{.4in} \phi(t) = \pi/4\, .
\ee
Thus, as expected the vacuum $|0\rangle$ evolves into the highly squeezed state along a direction that is an equal mixture of the $\hat q$ and $\hat p$ directions.

An interesting concept in quantum mechanics that has enjoyed a fair amount of recent interest is the {\it circuit complexity} of a pair of states.
Defined in an analogous way to classical complexity, the circuit complexity is roughly the minimum number of fundamental quantum gates required to transform a reference state to some target state.

As discussed in the introduction, there are several different methods of computing the circuit complexity between a reference and a target state, including the geometric approaches by Nielsen's \cite{NL1,NL2,NL3}. Moreover, based on the choice of cost functional for each of these approaches there are different measures\cite{MyersCC,MyersCC1,NL1}. In the main part of the paper we will focus on the circuit complexity using directly the wavefunction \cite{MyersCC}.

To begin our calculation of complexity of the inverted harmonic oscillator we first need to obtain the position-space wavefunction for the squeezed state $|\Psi(t)\rangle$
\be
\langle x|\Psi(t)\rangle = {\mathcal N} e^{-\frac{1}{2} \Omega(t) x^2}\, ,
\label{invertGaussian}
\ee
where ${\mathcal N}$ is a normalization factor and $\Omega(t)$ is the complex frequency
\be
\Omega(t) = \frac{k}{e^{2r} \sin^2 \phi + e^{-2r} \cos^2 \phi}\left(1 -i\sin(2\phi) \sinh(2r)\right)\ .
\label{invertGaussianFreq}
\ee
In the unsqueezed limit $r \rightarrow 0$ we obtain the unsqueezed ground state wavefunction with $\Omega(t) \approx k$.
In the highly squeezed limit, however, where $\phi \approx \pi/4$ and $r \gg 1$ we obtain a purely complex frequency $\Omega(t) \approx i$.

Taking the unsqueezed vacuum $\langle x |0\rangle$ as our reference state and the squeezed state $\langle x |\Psi(t)\rangle$ (\ref{invertGaussian}) as our target state, 
the geometric circuit complexity evaluates to be \cite{AB}:
\be
\label{InvertedComplexity1}
{\mathcal C}_1 &=& \frac{1}{2} \left[ \ln\left|\frac{\Omega(t)}{k}\right|+\tan^{-1} \left(\frac{{\rm Im}\ \Omega(t)}{{\rm Re}\Omega}\right)\right]\, ; \\
{\mathcal C}_2 &=& \frac{1}{2} \sqrt{\left(\ln\left|\frac{\Omega(t)}{k}\right|\right)^2 + \left(\tan^{-1} \left(\frac{{\rm Im}\ \Omega}{{\rm Re} \Omega}\right)\right)^2}\, ,
\label{InvertedComplexity2}
\ee
 where ${\mathcal C}_1,{\mathcal C}_2$ refer to the complexity calculated with different cost functionals, as we explain in more detail in Section \ref{sec:Complexity}.
For small amounts of squeezing $r \ll 1$ we have
\be
{\mathcal C}_1 \sim {\mathcal C}_2 \approx 0\, ,
\ee
as expected, since then the reference and target states are approximately the same.
For large amounts of squeezing $r \gg 1, \phi \approx \pi/4$ (corresponding to late times for the inverted harmonic oscillator), these expressions for the complexity (\ref{InvertedComplexity1},\ref{InvertedComplexity2}) become

\be
{\mathcal C}_1\sim {\mathcal C}_2 \approx  \frac{1}{2} \sqrt{\left(\tan^{-1} e^{2r}\right)^2} \approx \frac{\pi}{4}\, ,
\label{InvertedComplexityResult}
\ee
so that the complexity of a single mode vacuum squeezed state {\it saturates} at late times.
This is consistent with the expectation that the complexity for a quantum chaotic system saturates at some maximum complexity.

More generally, squeezed vacuum states are frequently used in quantum optics applications outside of the context of the inverted harmonic oscillator, so the results found here are of more general interest and applicability.
In this sense, we can take the general squeezed state (\ref{InvertedPsi}) -- and its gaussian form (\ref{invertGaussian}),(\ref{invertGaussianFreq}) -- as representing a generic squeezed vacuum state.
We can then easily determine the complexity of such a squeezed vacuum state from the expressions (\ref{InvertedComplexity1}),(\ref{InvertedComplexity2}).
In particular, note that if the squeezing angle is fixed to be $\phi \rightarrow n \frac{\pi}{2}$ for some integer $n$
then the complexity of the squeezed state (\ref{InvertedPsi}) (equivalently (\ref{invertGaussian})) 
does not saturate, but instead scales with the squeezing ${\mathcal C}_1 \sim r$ for large squeezing $r \gg 1$.

\section{Squeezed Cosmological Perturbations}
\label{sec:SqueezedCosmo}

Having explored the concepts of squeezing and complexity in a simple model of an inverted harmonic oscillator, we are now ready to apply these concepts to that of scalar cosmological perturbations.

We will consider a spatially flat Friedmann-Lemaitre-Robertson-Walker (FLRW) metric
\be
ds^2 = -dt^2 + a(t)^2 d\vec{x}^2 = a(\eta)^2 \left(-d\eta^2+d\vec{x}^2\right)\, .
\ee
On this background we will consider fluctuations of a scalar field $\varphi(x) = \varphi_0(t) + \delta\varphi(x)$ and the metric
\be
ds^2 = a(\eta)^2 \left(-(1+2\psi(x,\eta))d\eta^2 + (1-2\psi(x,\eta)) d\vec{x}^2\right)\, .
\ee
The perturbed action can be written in terms of the curvature perturbation ${\mathcal R} = \psi+\frac{H}{\dot \varphi_0} \delta \varphi$, where a dot denotes a derivative with respect to cosmic time $t$, and $H = \dot a/a$.
The action then takes the simple form \cite{Mukhanov}
\be
S = \frac{1}{2}\int dt\, d^3x\, a^3 \frac{\dot \phi^2}{H^2} \left[\dot {\mathcal R}^2 - \frac{1}{a^2} \left(\partial_i {\mathcal R}\right)^2\right]\, .
\ee
The action can be transformed into a form of that for a canonically normalized scalar field by use of the Mukhanov variable
$v \equiv z {\mathcal R}$
where $z \equiv a\, \sqrt{2\epsilon}$, with $\epsilon = -\dot H/H^2 = 1-{\mathcal H}'/{\mathcal H}^2$,
\be
S = \frac{1}{2} \int d\eta\, d^3x \left[v'^2 - (\partial_i v)^2 + \left(\frac{z'}{z}\right)^2 v^2 - 2 \frac{z'}{z} v' v\right]\, .
\label{CosmoAction}
\ee
Here a prime denotes a derivative with respect to conformal time and ${\mathcal H} = a'/a$.
This action represents perturbations of a free scalar field coupled to an external time-varying source. 
A virtually identical-looking expression can also be derived for tensor perturbations with the replacement $z'/z \rightarrow a'/a$, and our results will hold for these types of perturbations as well.
Usually the last term in (\ref{CosmoAction}) is removed by integration by parts\footnote{The last term in (\ref{CosmoAction}) can also be removed by an appropriate canonical transformation as discussed in \cite{Martin1}.}, giving rise to the action
\be
S = \frac{1}{2} \int d\eta\, d^3x \left[v'^2 - (\partial_i v)^2 + \frac{z''}{z}v^2\right]\, .
\label{CosmoHarmonicAction}
\ee
In this form, the time-varying source clearly leads to a time-dependent frequency, and this can cause the long-wavelength modes 
to appear as an inverted harmonic oscillator.
While we will be working instead with the action (\ref{CosmoAction}), the physics will nonetheless follow this intuition.

Promoting the perturbation to a quantum field and expanding into Fourier modes
\be
\hat v(\eta,\vec{x}) = \int \frac{d^3k}{(2\pi)^{3/2}} \hat v_{\vec{k}}(\eta)\, e^{i\vec{k}\cdot \vec{x}}\, ,
\ee

and defining the usual creation and annihilation operators\\

\be
\hat v_{\vec{k}} = \frac{1}{\sqrt{2k}} \left(\hat c_{\vec{k}} + \hat c_{-\vec{k}}^\dagger\right), \hspace{.2in}
\hat v_{\vec{k}}' = -i\frac{k}{2} \left(\hat c_{\vec{k}} - \hat c_{-\vec{k}}^\dagger\right)\, ,
\label{CosmoCreationOperators}
\ee
the Hamiltonian can be written as
\be
\hat H = \int d^3k\, \hat {\mathcal H}_{\vec{k}} = \int d^3k \left[k\left(\hat c_{\vec{k}} \hat c_{\vec{k}}^\dagger + \hat c_{-\vec{k}}^\dagger \hat c_{-\vec{k}}\right) 
    - i \frac{z'}{z} \left(\hat c_{\vec{k}} \hat c_{-\vec{k}} - \hat c_{\vec{k}}^\dagger \hat c_{-\vec{k}}^\dagger\right)\right]\, .
\label{CosmoH}
\ee
The first term in (\ref{CosmoH}) represents the usual free-particle Hamiltonian, while the second term describes the interaction between the quantum perturbation and the expanding background.
Notice that this last term is similar in form to the Hamiltonian (\ref{InvertH}) for the inverted harmonic oscillator from the last section, and indeed we will see that when the last term in the Hamiltonian dominates $z'/z \gg k$ the squeezing for the curvature perturbation will also grow.
The momentum structure of the Hamiltonian indicates that the interaction with the background leads to particle creation in pairs with opposite momenta. Because of this,  we are naturally led to consider our states as appearing in two-mode pairs $(\vec{k},-\vec{k})$.

As with the inverted harmonic oscillator, the unitary evolution ${\mathcal U_{\vec{k}}}$ of a state can be factorized into a parameterization of the form
\cite{Grishchuk,Albrecht}
\be
\hat {\mathcal U}_{\vec{k}} = \hat{\mathcal S}_{\vec{k}}(r_k,\phi_k) \hat{\mathcal R}_{\vec{k}}(\theta_k)\, ,
\ee
where $\hat{\mathcal R}_{\vec{k}}$ is the two-mode rotation operator
\be
\hat{\mathcal R}_{\vec{k}}(\theta_k) \equiv {\rm exp}\ \left[-i\theta_k(\eta) (\hat c_{\vec{k}} \hat c_{\vec{k}}^\dagger + \hat c_{-\vec{k}}^\dagger \hat c_{-\vec{k}})\right]
\ee
written in terms of the rotation angle parameter $\theta_k(\eta)$ and $\hat{\mathcal S}_{\vec{k}}$ is the two-mode squeeze operator
\be
\hat {\mathcal S}_{\vec{k}}(r_k,\phi_k) \equiv {\rm exp}\ \left[\frac{r_k(\eta)}{2} \left(e^{-2i\phi_k(\eta)} \hat c_{\vec{k}} \hat c_{-\vec{k}} - e^{2i\phi_k(\eta)} \hat c_{-\vec{k}}^\dagger \hat c_{\vec{k}}^\dagger\right)\right]
\ee
written in terms of the squeezing parameter $r_k(\eta)$ and squeezing angle $\phi_k(\eta)$.
As with the inverted harmonic oscillator, the rotation operator and rotation angle $\theta_k$ will not be important, so we will not include them in our subsequent analysis.
Also, since the squeezing equations of motion will only depend on the magnitude $k$ of the wavenumber $\vec{k}$, we have suppressed the vector notation on the subscripts of these parameters.

By recognizing that the interaction of the cosmological perturbation with the time-dependent scale factor leads to a time-dependent frequency for the canonically normalized harmonic oscillator (\ref{CosmoHarmonicAction}), the appearance of a squeezed state for cosmological perturbations is quite natural in the context of the previous section on the inverted harmonic oscillator.
The quantization of this parametric oscillator is then naturally described in the language of two-mode squeezed states \cite{Grishchuk,Albrecht,Martin1,Martin2}.

We will assume that at the initial time all of the modes of interest are well inside the horizon $k \gg |\eta|$ so that the system can be described by the free part of the Hamiltonian (\ref{CosmoH}). We then define the initial state (two-mode) vacuum with respect to the annihilation operator
\be
\hat c_{\vec{k}} |0\rangle_{\vec{k},-\vec{k}} = 0, \hspace{.2in} \forall\ \vec{k}\, .
\ee
The two-mode squeeze operator results in a two-mode squeezed vacuum state
\begin{equation}
    |\Psi_{sq}\rangle_{\vec{k},-\vec{k}} = \hat {\mathcal S}(r_k,\phi_k)_{\vec{k}} |0\rangle_{\vec{k}} = \frac{1}{\cosh r_k} \sum_{n=0}^{\infty} (-1)^n e^{-2 i n \phi_k} \tanh^n r_k\, |n_{\vec{k}}; n_{-\vec{k}}\rangle\, ,
    \label{psi1}
\end{equation}
where the two-mode excited state is
\be
|n_{\vec{k}}; n_{-\vec{k}}\rangle = \sum_{n=0}^\infty \frac{1}{n!} \left(\hat c_{\vec{k}}^\dagger\right)^n \left(\hat c_{-\vec{k}}^\dagger\right)^n\, |0\rangle_{\vec{k},-\vec{k}}\, .
\ee
The full wavefunction then consists of the product of the wavefunctions for each $\vec{k}$
\be
|\Psi\rangle = \otimes_{\vec{k}} |\Psi\rangle_{\vec{k},-\vec{k}}\, ,
\ee
though we will mostly just work with $|\Psi\rangle_{\vec{k},-\vec{k}}$.
The time evolution of the squeezing parameters $r_k(\eta),\phi_k(\eta)$ is determined by the Schr\"odinger equation
\be
i \frac{d}{d\eta} |\Psi_{sq}\rangle_{\vec{k},-\vec{k}} = \hat {\mathcal H}_{\vec{k},-\vec{k}} |\Psi_{sq}\rangle_{\vec{k},-\vec{k}}\, ,
\ee
and leads to the differential equations
\begin{eqnarray} 
\frac{dr_k}{d\eta} &=&
-\frac{z'}{z} \cos (2\phi_k)\, ;~ \nonumber \\
\frac{d\phi_k}{d\eta}&=& k +
\frac{z'}{z} \coth(2r_k) \sin (2\phi_k)\, .
\label{martin22}
\end{eqnarray}
Note that for a stationary background spacetime $z$ is constant, so there is no squeezing $r = 0$.

\subsection{Squeezing Solutions}

For a given background expansion $a(\eta)$, the squeezing equations (\ref{martin22}) can be solved for the squeezing parameters $r_k(\eta),\phi_k(\eta)$ (recall $z \equiv a\sqrt{2\epsilon}$).
Before we proceed to compute the circuit complexity for the states (\ref{psi1}),  let us explore the behavior of squeezing solutions for cosmological backgrounds. This will give us some insight into the behavior of squeezing due to the expansion of the Universe. The squeezing of cosmological perturbations has been studied previously \cite{Grishchuk,Albrecht}.

In general the equations (\ref{martin22}) must be solved numerically for a given cosmological background.
However, we can make progress with a qualitative understanding of the solutions by noting that in general the scale factor depends on some power of the conformal time 
\be
a(\eta) \sim \left(\frac{\eta}{\eta_0}\right)^\beta 
= \begin{cases}
-\frac{1}{H\eta} & \beta = -1,\ \mbox{de Sitter} \\
\frac{\eta}{\eta_0} & \beta = 1,\ \mbox{Radiation} \\
\left(\frac{\eta}{\eta_0}\right)^2 & \beta = 2,\ \mbox{Matter}
\end{cases}\, ,
\label{scalefactor}
\ee
where $\beta = 2/(1+3w)$ in terms of the equation of state $p/\rho = w$ of the cosmological fluid of the background expansion.
These different equations of state can arise, for example, as the behavior of the scalar field $\varphi$
on different potentials $V(\varphi) = V_0 \varphi^\gamma$.
Accordingly, this implies that the term $z'/z$ appearing in the squeezing equations of motion scales inversely proportional to $\eta$: $z'/z = \beta/\eta$. The equations of motion (\ref{martin22}) then become
\be
\frac{dr_k}{d\eta} &=&
-\frac{\beta}{\eta} \cos (2\phi_k)\, ;~ \nonumber \\
\frac{d\phi_k}{d\eta}&=& k +
\frac{\beta}{\eta} \coth(2r_k) \sin (2\phi_k)\, .
\label{squeezeQual}
\ee
Solutions to (\ref{squeezeQual}) depend whether the mode is super-horizon $k|\eta| \ll 1$ or sub-horizon $k|\eta| \gg1$ and whether the squeezing is small $r_k \ll 1$ or large $r_k \gg 1$.

Let's begin by considering the small squeezing, sub-horizon limit. The equations of motion in this limit take the form
\be
\frac{dr_k}{d\eta} &=&
-\frac{\beta}{\eta} \cos (2\phi_k)\, ;~ \nonumber \\
\frac{d\phi_k}{d\eta}&=& k +
\frac{\beta}{\eta} \frac{1}{2r_k} \sin (2\phi_k)\, ;
\ee
where we took the small $r_k$ limit of $\coth(2r_k)$.
These equations of motion have the solution
$r_k \sim \beta/(2k\eta) \ll 1$ and $\phi_k \sim -\pi/4$, indicating that in the small squeezing, sub-horizon limit the squeezing stays small with fixed squeezing angle.
A similar analysis of the small squeezing, super-horizon limit has an approximate solution $r_k \sim |\beta \ln(k\eta)|$, $\phi_k \sim -\pi/2$. However since $k |\eta| \ll 1$ for super-horizon modes, this indicates there is tension 
with having a super-horizon mode with small squeezing, so we should instead consider super-horizon modes with large squeezing, for which the squeezing equations of motion take the form
\be
\frac{dr_k}{d\eta} &=&
-\frac{\beta}{\eta} \cos (2\phi_k)\, ;~ \nonumber \\
\frac{d\phi_k}{d\eta}&\approx& 
\frac{\beta}{\eta} \sin (2\phi_k)\, .
\label{LargeSqueezeEOMQualitative}
\ee
Here, we indeed see that solutions self-consistently take the form
$r_k \sim |\beta \ln(k\eta)|$, $\phi_k \sim -\pi/2$ for $k |\eta| \ll 1$.
Thus, we have learned that an initially small squeezing inside the horizon remains small until it exits the horizon, after which it begins to grow and becomes much larger than one. Note that since the squeezing scales as the log of the conformal time on super-horizon scales then it also is proportional to the number of e-folds for the mode $k$ since horizon exit $r_k \sim \ln a(\eta)/a_{exit} \equiv N_e^{(k)}$.

\begin{figure}[t]
\centering \includegraphics[width=.65\textwidth]{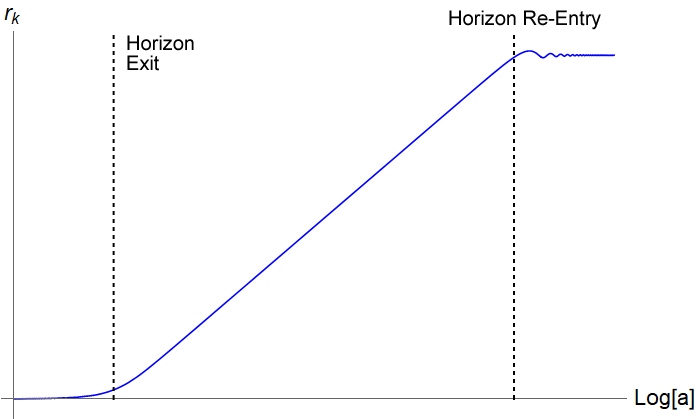}
\caption{In this qualitative plot, we follow the growth of the squeezing parameter $r_k$ as a function of the scale factor $a$ for fixed $k$ as it starts small inside the horizon, then grows larger than one after horizon exit, then ``freezes out'' upon horizon re-entry with a decaying oscillation, as described in the text. Notice that while outside of the horizon the squeezing parameter grows as the number of e-folds spent super-horizon $r_k \sim \log a \sim N_e^{(k)}$.}
\label{fig:QualSqueezing}
\end{figure}

Finally, we consider a mode which is highly squeezed but re-enters the horizon at some later time. This is what would happen, for example, for modes that exit the horizon during inflation, becoming highly squeezed then re-enter the horizon after the end of inflation during a radiation- or matter-dominated stage of expansion.
In this case the squeeze equation of motion for $\phi_k$ becomes
\be
\frac{d\phi_k}{d\eta} \approx k\, ,
\ee
so that the squeezing angle is no longer fixed but is instead running $\phi_k \sim \phi_k^{(0)} + k \eta$. Examining the corresponding equation for the squeezing parameter
\be
\frac{dr_k}{d\eta} \approx \frac{\beta}{\eta} \cos\left(2\phi_k^{(0)} + 2k \eta\right)\, ,
\label{largeSqueezeSubH}
\ee
we see that the running $\phi_k$ will cause $\cos(2\phi_k)$ to oscillate between positive and negative values, shutting off growth of $r_k$.
Indeed, an approximate solution to (\ref{largeSqueezeSubH}) is a damped oscillation
\\
%
\be
r_k \sim r_k^{(0)} + \frac{\beta}{2k\eta} \sin\left(2\phi_k^{(0)} + 2k \eta\right)\, .
\ee
Thus, when highly squeezed mode re-enters the horizon it ``freezes in'' to the value of the squeezing at horizon-crossing, with a decaying oscillation about that value.
A plot illustrating these qualitative features -- no squeezing growth on sub-horizon scales, squeezing growth on super-horizon scales, and freeze-out of squeezing upon horizon re-entry -- is shown in Figure \ref{fig:QualSqueezing}. Below we will consider some exact and numerical solutions to the full squeezing equations of motion (\ref{martin22}), and we will see precisely these features.

With a general qualitative understanding of the behavior of squeezing solutions in hand, now let's explore some exact and numerical solutions to (\ref{martin22}) for some specific cosmological backgrounds.
The simplest solution is that of an exponentially expanding de Sitter background, for which $a(\eta) = -1/(H\eta)$ for $-\infty < \eta < 0$, so that $z'/z = -1/\eta$.
An exact solution for a de Sitter background is known\footnote{Note that there is a typo involving a factor of $1/2$ in the solution for $\phi_k$ in the solution of \cite{Albrecht}.} \cite{Albrecht}
\be
r_k &=& -\sinh^{-1} \left(\frac{1}{2k\eta}\right)\, ; \nonumber \\
\phi_k &=& -\frac{\pi}{4} + \frac{1}{2} \tanh^{-1} \left(\frac{1}{2k\eta}\right)\, .
\label{exactdSSqueezing}
\ee
At early times $k |\eta| \gg 1,$ the modes are inside the horizon, and we have vanishing squeezing $r_k \approx -\frac{1}{2k\eta} \ll 1,$ and an approximately constant squeezing angle $\phi_k \approx -\pi/4$, as already discussed in our qualitative analysis. At late times $k |\eta| \ll 1$ the modes are outside the horizon; from the action (\ref{CosmoHarmonicAction}) in which the modes appear as a harmonic oscillator with a time-dependent frequency, the external frequency due to the expansion of the Universe dominates and the action takes the form of an inverted harmonic oscillator.
Thus, we expect in this regime that the squeezing will grow with time, as did the inverted harmonic oscillator from the previous section.
Indeed, in this limit the solution (\ref{exactdSSqueezing}) gives a growing squeezing parameter $r_k \approx |\ln(-k\eta)| \sim \ln(a) \gg 1$ as $k |\eta| \gg 1$ and constant squeezing angle $\phi_k \approx -\pi/2$, again in excellent agreement with our qualitative analysis. Since the squeezing parameter grows with the log of the scale factor it is proportional to the number of e-folds of de Sitter expansion since horizon exit $r_k \sim N_e^{(k)}$, a feature we saw was true more generally for other expanding backgrounds.

\begin{figure}[t]
\centering\includegraphics[width=.48\textwidth]{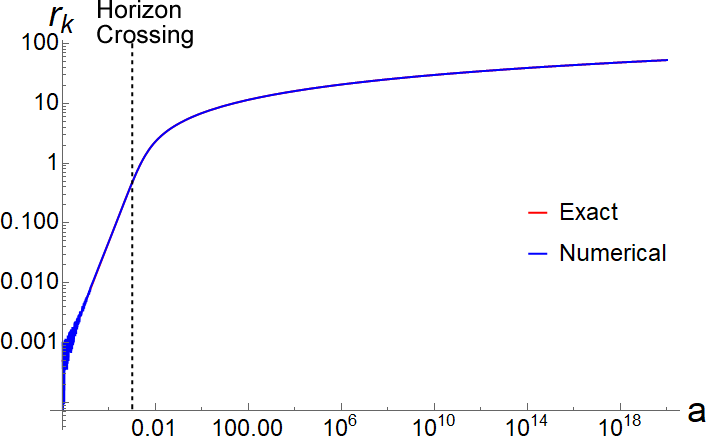} \hspace{.2in} \includegraphics[width=.48\textwidth]{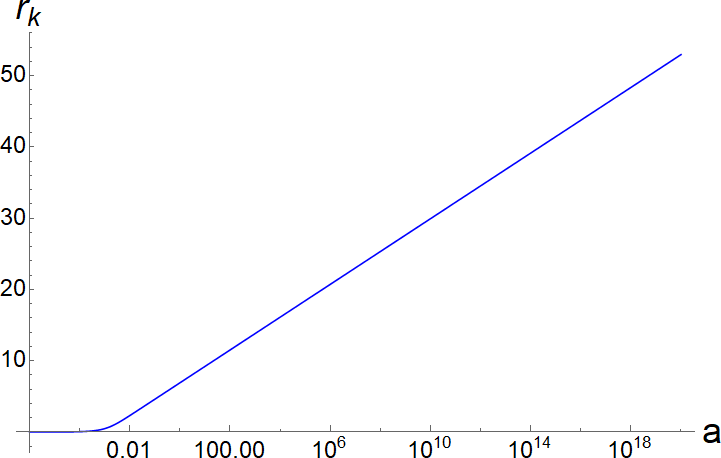}
\caption{(Left) The squeezing parameter $r_k$ as a function of the scale factor $a$ for de Sitter space for the exact solution (\ref{exactdSSqueezing}) and numerical solutions to the squeezing equations (\ref{martin22}) for $k = 0.001$ in units of $\eta_0$, defined by $a(\eta_0) = 1$ (color online). 
The squeezing parameter grows appreciably -- and logarithmically -- only on super-horizon scales $k < 1/|\eta|$. (Right) The same graph shown with a linear scale for $r_k$ demonstrates that the growth on super-horizon scales is proportional to the number of e-folds of expansion since mode $k$ exited the horizon $r_k \sim N_e^{(k)}$.}
\label{fig:dSSqueezing}
\end{figure}

\begin{figure}[t]
\centering\includegraphics[width=.47\textwidth]{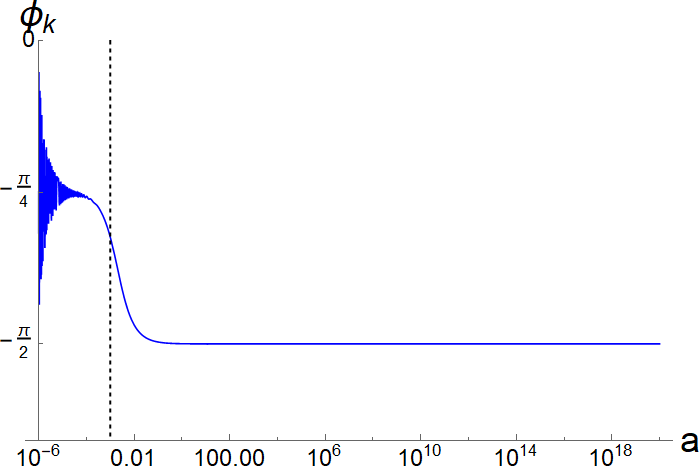} \hspace{.1in} \includegraphics[width=.47\textwidth]{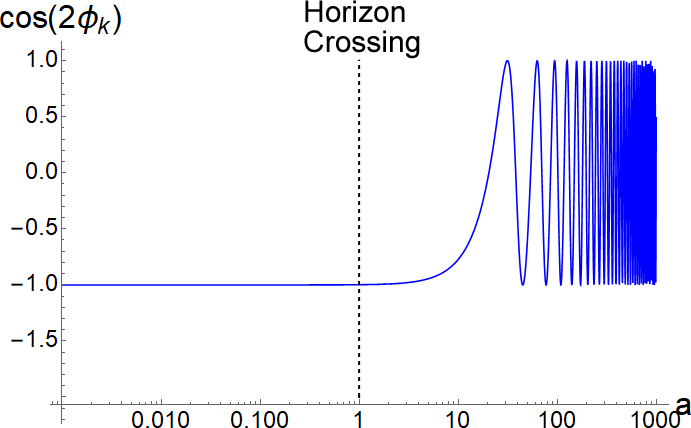}
\caption{(Left) The squeezing angle $\phi_k$ for a dS background (for the same $k$ as Figure \ref{fig:dSSqueezing}) oscillates around  $\phi_k = -\pi/4$ when the mode is inside the horizon, and then transitions to $\phi = -\pi/2$ after the mode exits the horizon, in accordance with our qualitative results from the text and the exact solution (\ref{exactdSSqueezing}).
(Right) The squeezing angle for a radiation-dominated background with $k = 0.1$ (again in units of $\eta_0$), plotted as $\cos(2\phi_k)$. Notice that at early times while the mode is super-horizon we have $\phi_k \approx -\pi/2$, while after the mode re-enters the horizon we have $\phi_k \sim k \eta$ increasing with time leading to oscillations in $\cos(2\phi_k)$ which cuts off further growth in $r_k$, in agreement with our qualitative analysis in the text.}
\label{fig:SqueezingAngle}
\end{figure}

Based on this analysis, we see that the vacuum state will remain un-squeezed while modes are inside the horizon, while squeezing will begin to grow appreciably once modes exit the horizon. Since in dS space modes that begin inside the horizon eventually exit the horizon due to the expansion of the Universe, we expect that an initially un-squeezed vacuum state for a mode $\vec{k}$ will become increasingly squeezed as time evolves in a de Sitter Universe.
Indeed, we see precisely this behavior in the analytic solution (\ref{exactdSSqueezing}) as well as numerical solutions to the squeezing equations (\ref{martin22}), as shown in Figures \ref{fig:dSSqueezing} and \ref{fig:SqueezingAngle}.

\begin{figure}[t!]
\centering\includegraphics[width=.47\textwidth]{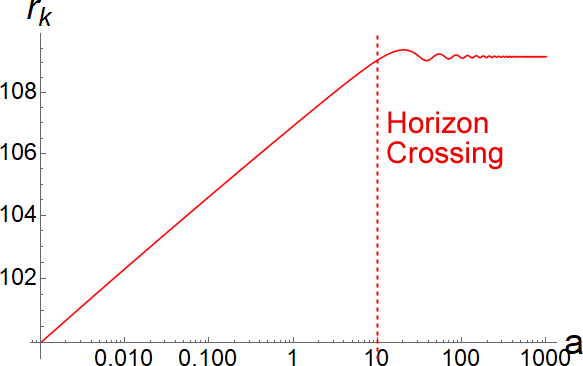}\hspace{.1in} \includegraphics[width=.47\textwidth]{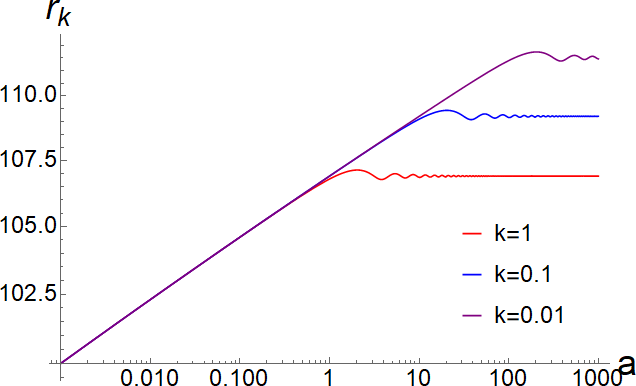}
\caption{(Left) The squeezing parameter $r_k$ for a radiation background with $k = 0.1$ in units of $\eta_0$ is plotted against the scale factor $a$. Since modes start outside the horizon in a radiation background, the squeezing is large and growing at early times. Once the mode re-enters the horizon, however, the squeezing ``freezes in'' with a damped oscillation about the value at horizon crossing. (Right) Different wavenumbers (again in units of $\eta_0$) lead to different times of horizon re-entry, and thus different ``freeze in'' values of the squeezing.}
\label{fig:RadiationSqueezing}
\end{figure}

For a cosmological background dominated by radiation we have $a(\eta) = \eta/\eta_0$, so that $z'/z = 1/\eta$, where now $\eta > 0$. 
This background could arise in the presence of a scalar field due to the oscillation of the homogeneous scalar field condensate about a minimum, such as for example during reheating after the end of inflation.
Interestingly, a slight modification to the signs of the exact de Sitter solution (\ref{exactdSSqueezing}) leads to an exact solution for radiation as well:
\be
r_k &=& \sinh^{-1} \left(\frac{1}{2k\eta}\right)\, ; \nonumber\\
\phi_k &=& -\frac{\pi}{4} + \frac{1}{2} \tanh^{-1}\left(\frac{1}{2k\eta}\right)\, .
\label{exactRadiationSqueezing}
\ee
Unlike the de Sitter case, however, at sufficiently early times $\eta \rightarrow 0$ a mode will start outside the horizon $k \eta \ll 1$, then re-enter the horizon later. This exact solution (\ref{exactRadiationSqueezing}), then, represents the {\it decaying} solution; we also expect there to be a growing mode solution as well.
Indeed, from the qualitative discussion above, we expect that the squeezing of the mode will continue to grow while outside of the horizon, then ``freeze in'' when the mode re-enters the horizon.
In Figure \ref{fig:RadiationSqueezing} we see precisely this behavior, where the squeezing parameter is plotted for several different magnitudes of the wavenumber $k$.
In Figure \ref{fig:SqueezingAngle} we see that the behavior of the squeezing angle before and after horizon re-entry matches our qualitative analysis from above, where $\phi_k \approx -\pi/2$ outside the horizon, and $\phi \sim k\eta$ after horizon re-entry.

\begin{figure}[t]
\centering\includegraphics[width=.60\textwidth]{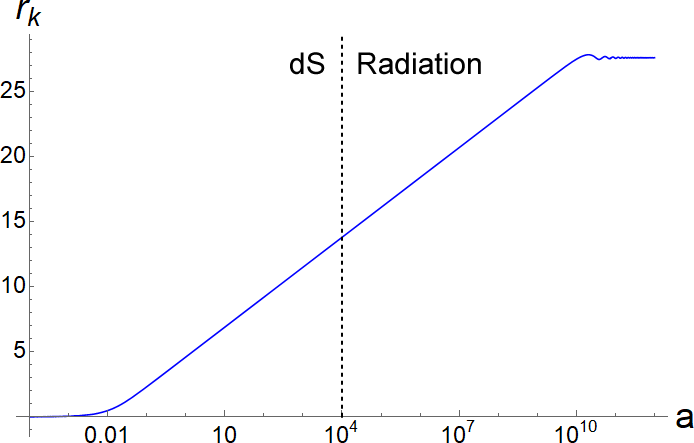}
\caption{The squeezing parameter $r_k$ for a cosmological background consisting of de Sitter followed by radiation shows the features already seen in the de Sitter and radiation plots separately ($k=0.01$ in units of $\eta_0$). The squeezing, initially small, grows upon horizon exit and continues growing through the transition to radiation. Eventually the mode re-enters the horizon during the radiation era and ``freezes out'' at its value at horizon crossing.}
\label{fig:UniverseSqueezing}
\end{figure}

\begin{figure}[h]
\centering\includegraphics[width=.95\textwidth]{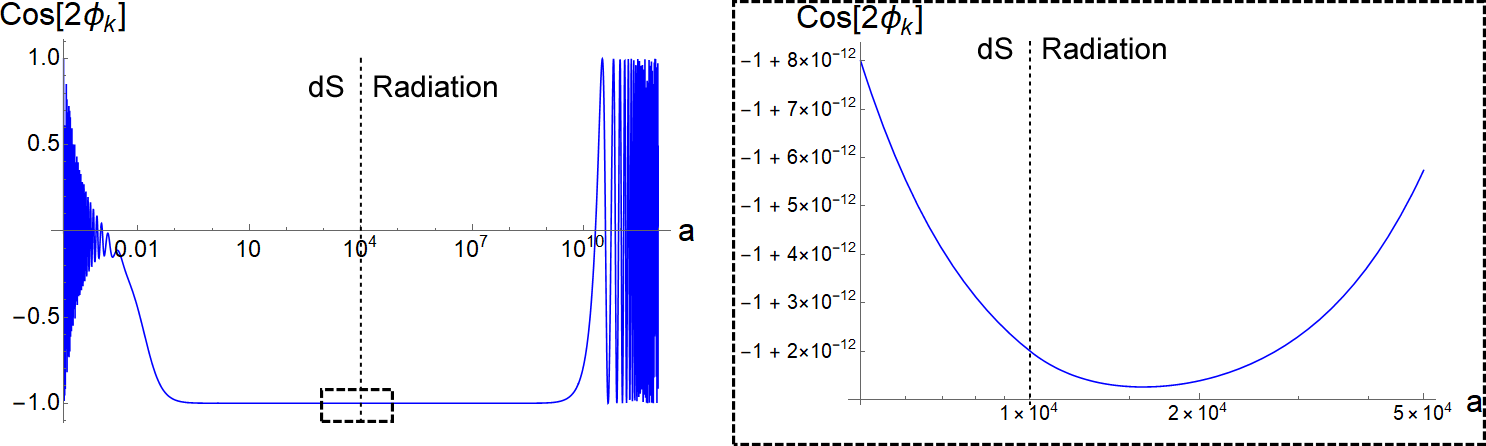}
\caption{(Left) The squeezing angle $\cos(2\phi_k)$ for the solution shown in Figure \ref{fig:UniverseSqueezing} shown as a function of the scale factor $a$ for a dS expansion followed by a transition to radiation shows how the squeezing angle freezes out to $\phi_k \approx -\pi/2$ when outside the horizon, and grows when it re-enters the horizon. (Right) The inset shows a zoomed in region of the transition between dS and radiation. Notice that the squeezing angle reaches a minimum some time after the transition, then begins to slowly grow again. This feature will be important in our understanding of the complexity in the next section.}
\label{fig:UniverseSqueezedAngle}
\end{figure}

Finally, let's consider a slightly more realistic background expansion that transitions from de Sitter at early times to radiation at late times. This can be viewed as a simple model of early Universe inflation followed by a period of scalar field reheating.
For this expansion history we expect modes starting inside the horizon to eventually exit the horizon, with corresponding growth in squeezing.
At the transition to radiation we don't expect to see any change in the growth of the squeezing parameter $r_k$;
however, at some point following this transition the mode will re-enter the horizon and the squeezing will ``freeze in''.
Figure \ref{fig:UniverseSqueezing} illustrates precisely this behavior.
The squeezing angle also illustrates similar behavior as we saw with the dS and radiation backgrounds separately, seen in Figure \ref{fig:UniverseSqueezedAngle}.
Interestingly, if we zoom in on the transition between dS and radiation for the squeezing angle we see that the squeezing angle reaches a minimum after some time after the actual transition; this feature will be important for our understanding of complexity for these combined backgrounds.

\section{Complexity for cosmological Squeezed States}

\label{sec:Complexity}

In the previous section, we saw that it is natural to describe the evolution of scalar cosmological perturbations as a two-mode squeezed vacuum state. We developed a qualitative understanding of the behavior of the squeezed solutions both inside and outside of the horizon, finding that in general, the corresponding quantized harmonic oscillator becomes inverted when modes become super-horizon, leading to squeezing in a similar way as we saw in Section \ref{sec:Inverted}.
We verified this qualitative reasoning with an exact solution in the case of a de Sitter expanding background, as well as numerical solutions for several other expanding backgrounds.

We are now ready to consider the complexity of the squeezed cosmological perturbations.
As discussed in Appendix \ref{app:Complexity}, we will compute the {\it circuit complexity} of a target state relative to a chosen reference state.
A natural reference state for our cosmological perturbations is that of the two-mode vacuum state $|0\rangle_{\vec{k},-\vec{k}}$, while our target state will be the squeezed two-mode vacuum state $|\Psi_{sq}\rangle_{\vec{k},-\vec{k}}$ in (\ref{psi1}).
In order to utilize the formalism of \cite{MyersCC}, we will need to express the reference and target states as gaussian wavefunctions. 
We will first define a set of auxiliary ``position'' and ``momentum'' variables
\be
\hat q_{\vec{k}} \equiv  \frac{1}{\sqrt{2k}} \left(\hat c_{\vec{k}}^\dagger + \hat c_{\vec{k}}\right), \hspace{.2in}
\hat p_{\vec{k}} \equiv  i\sqrt{\frac{k}{2}}\left(\hat c_{\vec{k}}^\dagger - \hat c_{\vec{k}}\right)\, ,
\ee
which are conjugate variables $[\hat q_{\vec{k}},\hat p_{\vec{k}'}] = i \delta^3(\vec{k}-\vec{k}')$.
Notice that the main difference between the ``position'' $\hat q_{\vec{k}}$ and the Fourier mode $\hat v_{\vec{k}}$ given in (\ref{CosmoCreationOperators}) is that the former is defined with respect to a raising operator of $\vec{k}$ instead of $-\vec{k}$.

The two-mode vacuum state's wavefunction, defined as $\hat c_{\vec{k}} |0\rangle_{\vec{k},-\vec{k}} = 0$, has the usual gaussian form
\be
\psi_R(q_{\vec{k}},q_{-\vec{k}})= \langle q_{\vec{k}},q_{-\vec{k}} | 0\rangle_{\vec{k},-\vec{k}} = \left(\frac{k}{\pi}\right)^{1/4}\, e^{-\frac{k}{2} (q_{\vec{k}}^2 + q_{-\vec{k}}^2)}\, .
\label{reference}
\ee
To calculate the wavefunction corresponding to the squeezed state (\ref{psi1}) we note that the following combination annihilates $|\Psi_{sq}\rangle_{\vec{k},-\vec{k}}$
\be
\left(\cosh r_k\ \hat c_{\vec{k}} + e^{-2i\phi_k} \sinh r_k\ \hat c_{-\vec{k}}^\dagger\right) |\Psi_{sq}\rangle_{\vec{k},-\vec{k}} = 0\, .
\ee
Using this we can calculate the ``position-space'' form of the wavefunction \cite{Martin2}
\begin{equation} \label{state1}
\Psi_{sq} (q_{\vec{k}}, q_{-\vec{k}})= \langle q_{\vec{k}},q_{-\vec{k}}|\Psi_{sq}\rangle_{\vec{k}} =  \frac{e^{A(q_{\vec{k}}^2+q_{-\vec{k}}^2)-B q_{\vec{k}} q_{-\vec{k}}}}{\cosh r_k \sqrt{\pi} \sqrt{ 1- e^{-4 i \phi_k} \tanh^2 r_k}}\, ,
\end{equation}
where the coefficients $A$ and $B$ are functions of the squeezing parameter $r_k$ and squeezing angle $\phi_k$
\begin{equation}
    A= \frac{k}{2} \left( \frac{e^{-4 i \phi_k} \tanh^2 r_k +1}{e^{-4 i \phi_k} \tanh^2 r_k -1} \right)\, ,\hspace{.2in} B= 2k \left( \frac{e^{-2 i \phi_k} \tanh r_k }{e^{-4 i \phi_k} \tanh^2 r_k -1}\right)\, .
    \label{ABSqueezed}
\end{equation}

As discussed in Appendix \ref{app:Complexity},
we will focus our study of complexity by directly working with the wavefunction using the approach of Nielsen \cite{NL1,NL2,NL3}, which we will call {\it circuit complexity}, though we do briefly investigate circuit complexity using covariance matrix method in Appendix \ref{app:Covariance}.
 Even selecting this general approach, however, does not eliminate all possible ambiguity in the computation of complexity, since there are different measures of complexity depending on different choices for the ``cost function.''
In particular, the complexity for two simple choices of cost functions -- ``linear'' weighting ${\mathcal C}_1$ and ``geodesic'' weighting ${\mathcal C}_2$ -- can easily be computed from the vacuum reference state (\ref{reference}) and squeezed target state (\ref{state1}) (see Appendix \ref{app:Complexity} for details)
\begin{eqnarray}
\label{complexity1}
    {\mathcal C}_1(k) &=&\frac{1}{2}  \left (\ln \left|\frac{\Omega_{\vec{k}}}{\omega_{\vec{k}}}\right|+  \ln \left|\frac{\Omega_{-\vec{k}}}{\omega_{-\vec{k}}}\right|+ \tan^{-1} \frac{\text{Im}\ \Omega_{\vec{k}}}{\text{Re}\ \Omega_{\vec{k}}} + \tan^{-1} \frac{\text{Im}\ \Omega_{-\vec{k}}}{\text{Re}\ \Omega_{-\vec{k}}} \right)\, ; \\ \cr
    {\mathcal C}_2(k) &=&\frac{1}{2} \sqrt{ \left(\ln \left|\frac{\Omega_{\vec{k}}}{\omega_{\vec{k}}}\right| \right)^2+ \left(\ln \left|\frac{\Omega_{-\vec{k}}}{\omega_{-\vec{k}}}\right| \right)^2+ \left(\tan^{-1} \frac{\text{Im}\ \Omega_{\vec{k}}}{\text{Re}\ \Omega_{\vec{k}}}\right)^2+ \left(\tan^{-1} \frac{\text{Im}\ \Omega_{-\vec{k}}}{\text{Re}\ \Omega_{-\vec{k}}}\right)^2},
\label{complexity2}
\end{eqnarray}
where $\Omega_{\vec{k}}=-2 A+B$, $\Omega_{-\vec{k}}=-2A-B$, and $\omega_{\vec{k}} = \omega_{-\vec{k}} = k/2$. The inverse tangent term in the above expression is necessary when the frequency is complex, see \cite{AB}.
We will see that the qualitative results for our squeezed states are essentially identical for these two measures (they only differ by a multiplicative factor) so we will have confidence in the genericity of our results\footnote{There will be some differences when compared against the circuit complexity computed using the covariance matrix; see Appendix \ref{app:Covariance}. However, as previously noted, we expect the latter to be less sensitive to detailed features of the wavefunction.}.

Using (\ref{ABSqueezed}) in (\ref{complexity1}),(\ref{complexity2}) we can obtain simple expressions for the two measures of complexity for the general two-mode squeezed vacuum state relative to the un-squeezed vacuum
\be
\label{complexSqueeze1}
    {\mathcal C}_1(k) &=&\left| \ln \left|\frac{1+e^{-2i\phi_k} \tanh r_k}{1-e^{-2i\phi_k}\tanh r_k}\right|\right|+ |\tan^{-1} \left(2\sin 2 \phi_k \sinh r_k \cosh r_k \right)|\, ; \\
    {\mathcal C}_2(k) &=& \frac{1}{\sqrt{2}} \sqrt{\left(\ln \left|\frac{1+e^{-2i\phi_k} \tanh r_k}{1-e^{-2i\phi_k}\tanh r_k}\right|\right)^2 + \left(\tan^{-1} \left(2\sin 2 \phi_k \sinh r_k \cosh r_k \right)\right)^2}\, .
\label{complexSqueeze2}
\ee
For large amounts of complexity $r_k \gg 1$ the last term is bounded by $\pi/2$, so the two measures of complexity are approximately equal to each other up to a multiplicative factor ${\mathcal C}_1 \approx \sqrt{2}\, {\mathcal C}_2$;
further, on super-horizon scales we expect the squeezing angle to take the value $\phi_k \rightarrow -\pi/2$, so the complexities (\ref{complexSqueeze1}),(\ref{complexSqueeze2}) simplify to be  simply proportional to the squeezing parameter
\be
{\mathcal C}_1(k) \approx \sqrt{2}\, {\mathcal C}_2(k) \approx \left|\ln\left(\frac{1-\tanh r_k}{1+\tanh r_k}\right)\right| \approx r_k \approx \ln a/a_{exit} = N_e^{(k)}\, ,
\label{complexityQualitative}
\ee
and therefore also proportional to the number of e-folds the mode $k$ has been super-horizon, as discussed below (\ref{LargeSqueezeEOMQualitative}). 
Since the expressions (\ref{complexSqueeze1}),(\ref{complexSqueeze2}) are functionally similar, we will focus our analysis on ${\mathcal C}_2$ without loss of generality.
It is interesting to note that that, in contrast to the inverted harmonic oscillator, we have found that the complexity grows with time, rather than saturating.
This appears to be due to the fact that while Hamiltonian for the inverted harmonic oscillator was time-independent, the term $z'/z$ in the Hamiltonian (\ref{CosmoH}) for cosmological perturbations is time-dependent, thus leading to growing complexity with time.
Note that (\ref{complexityQualitative}) implies that the rate of change of the complexity (with respect to cosmic time $t$) when the mode is super-horizon is given by the Hubble expansion rate
\be
\frac{d \mbox{Complexity}}{dt} \approx H\, .\ee

\subsection{Complexity in Expanding Backgrounds}

\begin{figure}[t] 
\centering
\includegraphics[width=.45\textwidth]{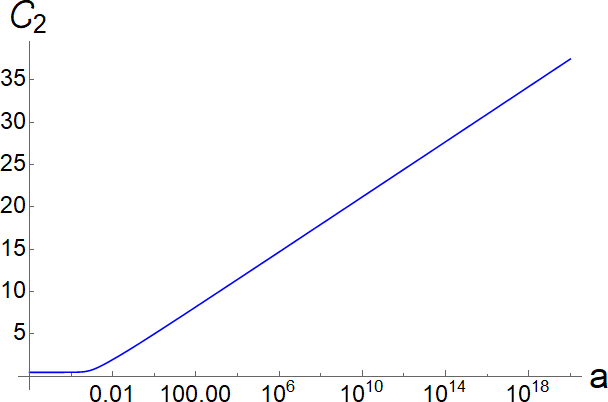}\hspace{.1in} \includegraphics[width=.45\textwidth]{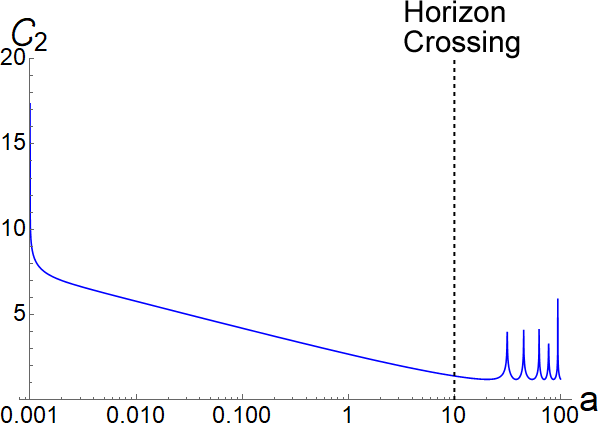}
\caption{(Left) The complexity ${\mathcal C}_2$ for a  cosmological perturbation in a dS cosmological background relative to the ground state reference demonstrates that the complexity remains small while the mode is within the horizon, then grows linearly with the $\log$ of the scale factor after exiting the horizon. ($k=0.001$ as in Figure \ref{fig:dSSqueezing}) (Right) The complexity for radiation illustrates a different pattern in which the complexity decreases from its starting value even while outside the horizon, due to the increasing squeezing angle $\phi_k$ for a radiation background as seen previously. This increasing squeezing angle leads to a decreasing complexity. After the mode re-enters the horizon for a radiation background it begins to oscillate, freezing in the complexity about which it subsequently oscillates. ($k=0.1$ as in Figure \ref{fig:RadiationSqueezing})}
\label{fig:Complexity}
\end{figure}

It is now a simple matter to insert the time-dependent solutions for the squeezing parameter and angle $r_k,\phi_k$ due to the expansion of the Universe from the previous section into (\ref{complexSqueeze2}) to see the time dependence of complexity for scalar cosmological perturbations.
Before we insert the numerical solutions, however, we can use our exact solutions for a dS expanding background (\ref{exactdSSqueezing})
to obtain analytic expressions for the complexity
(\ref{complexSqueeze2})
\be
{\mathcal C}_2(k) &=& \frac{1}{\sqrt{2}} \sqrt{\left(\log\left(\frac{(-2k\eta)} {\sqrt{4+(2k\eta)^2}}\right)\right)^2 + \left(\tan^{-1}\left(\frac{1}{-k\eta}\right)\right)^2} \\
&=& 
\begin{cases}
\frac{1}{\sqrt{2}} \sqrt{\frac{1}{(2k\eta)^2} + \left(\tan^{-1} \frac{1}{-k\eta}\right)^2} \approx \sqrt{\frac{5}{2}}\frac{1}{-2k\eta} & \mbox{ for $k\eta \gg 1$ (sub-horizon)} \cr
\frac{1}{\sqrt{2}} \sqrt{\left(\log(-k\eta)\right)^2 + \left(\frac{\pi}{2}\right)^2} \approx \frac{1}{\sqrt{2}} \left|\log(-k\eta)\right|\sim \frac{1}{\sqrt{2}} N_e^{(k)} & \mbox{ for $k\eta \ll 1$ (super-horizon)}
\end{cases}
\label{ComplexitydS}
\ee
where we note in the last line that the complexity scales like the number of e-folds for mode $k$ (as could be expected from (\ref{complexityQualitative})).
More generally, we can insert the numerical solutions for a dS background for the squeezing parameter and angle from the previous section into (\ref{complexSqueeze2}).
Figure \ref{fig:Complexity} shows that, as expected from our qualitative and exact analysis, when the mode is inside the horizon the complexity ${\mathcal C}_2$ is small, while when the mode exits the horizon the complexity quickly grows linearly with the $\log$ of the scale factor, and thus is proportional to the number of e-folds.

The linear growth of the complexity on super-horizon scales resembles the growth of complexity for other chaotic quantum systems \cite{qchaos1}, reflecting the fact that on super-horizon scales the Hamiltonian (\ref{CosmoH}) acts like an inverted harmonic oscillator.
As discussed in \cite{qchaos1}, we can extract information about quantum chaos \footnote{A concrete proof of chaos will require further tests by using other diagnostics of chaos. For example interested readers are referred to \cite{Kudler-Flam} and the references there in.} such as the scrambling time and Lyapunov exponent from the complexity.
Based on the analysis of \cite{qchaos1},
the scrambling time scale should be set by the time of horizon exit, and the Lyapunov exponent is set by the slope of the linear part of the complexity, which from (\ref{ComplexitydS}) is ${\mathcal O}(1)$. 

\begin{figure}[t] 
\centering
\includegraphics[width=.60\textwidth]{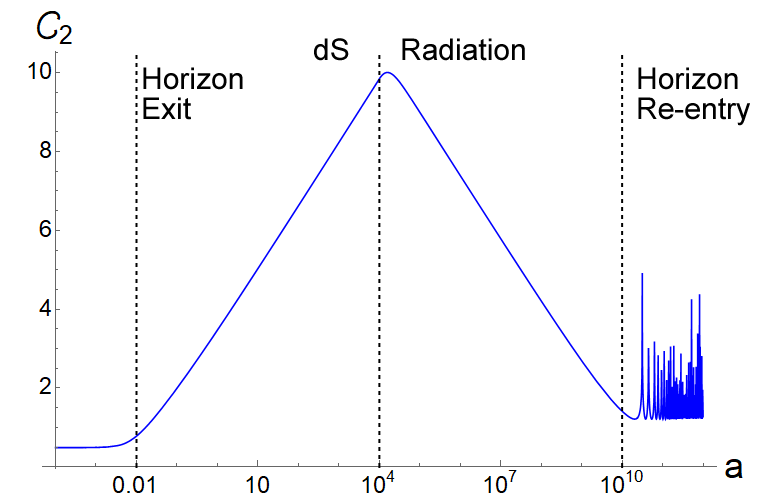}
\caption{The complexity ${\mathcal C}_2$ for the squeezing solution of Figures \ref{fig:UniverseSqueezing} and \ref{fig:UniverseSqueezedAngle}, namely a background that transitions from dS to radiation, initially grows on super-horizon scales during the dS phase, but decreases on super-horizon scales during the radiation phase, similar to that seen for pure radiation in Figure \ref{fig:Complexity}. After horizon re-entry, the complexity ``freezes in'' and oscillates due to the rapid evolution of the squeezing angle $\phi_k \sim k\eta$ on sub-horizon scales. The slight mismatch between the transition between dS and radiation and the peak of the complexity is due to the offset minimum in the squeezing angle $\phi_k$ after the transition, as seen in Figure \ref{fig:UniverseSqueezedAngle}.}
\label{fig:UniverseComplexity}
\end{figure}

Also in Figure \ref{fig:Complexity} we show the evolution of the complexity for a radiation background.
Contrary to the dS case, the complexity does not grow on super-horizon scales for a radiation background.
At first glance this seems puzzling, since the squeezing $r_k$ continues to grow on super-horizon scales, as seen in Figure \ref{fig:RadiationSqueezing}.
However, as seen in Figure \ref{fig:SqueezingAngle} (and in the detailed zoom of Figure \ref{fig:UniverseSqueezedAngle})
the squeezing angle $\phi_k$ increases during the radiation era $\phi_k \sim -\pi/2 + k\eta$, driving the complexity to lower values until horizon crossing. After horizon crossing the squeezing angle is now dominated by the sub-horizon contribution $\phi_k \sim k\eta$, leading to oscillations in the complexity through $e^{-2i\phi_k}$.
Thus, we see that unlike entropy, the circuit complexity of a mode can decrease.

Naturally, the evolution of the complexity for the simple model of the Universe consisting of a period of dS expansion followed by radiation is the concatenation of these two behaviors, as seen in Figure \ref{fig:UniverseComplexity}. As noted earlier, during the de Sitter era 
the complexity starts at close to zero since the mode is approximately that of the unsqueezed vacuum, and is nearly constant until the mode exits the horizon.
After horizon exit the complexity continues to grow as long as the Universe is accelerating.
During this period the linear growth of the complexity for super-horizon modes during the de sitter era resembles quantum chaos.
This scenario changes quite dramatically almost immediately after entering into the radiation regime. During this period the Universe de-complexifies and eventually after the mode re-enters the horizon the complexity ``freezes in'' at a value higher than the initial complexity before horizon exit.

Finally, we note that one can easily extend our analysis of complexity for all modes
\begin{equation}
    {\mathcal C}^{(\rm tot)}= \sum_k {\mathcal C}_2(k). 
    \label{TotalComplexity}
\end{equation}
As we have seen, a vacuum state that starts inside the horizon remains unsqueezed until it exits the horizon, with correspondingly small complexity. This means that ultra-high energy modes $k \eta \gg 1$ that don't exit the horizon before the transition to radiation will essentially not contribute at all to the total complexity of the Universe in this model, providing an effective UV cutoff to the complexity sum (\ref{TotalComplexity}).
The complexity is instead dominated by the first modes that exit the horizon, since they accumulate the largest amount of e-folds while super-horizon.
It would be interesting to carefully calculate the total complexity of the Universe for a more realistic background evolution for a future work.

\section{Discussion}
Quantum information theory is helping to shape our understanding about fundamental properties of nature, and quantum complexity plays a major role. In this paper we have applied Nielsen's geometric approach to compute the complexity of the Universe; specifically, we  computed the complexity of scalar cosmological perturbations by taking our reference state as the unsqueezed ground state and our target state as the squeezed vacuum state representing the evolution of cosmological perturbations.

This approach gives us a new perspective in which to examine the history of the Universe. We found that the complexity during dS expansion grows linearly with the number of e-folds for super-horizon modes, with the rate of change of complexity given by the dS Hubble expansion rate.
This linear growth suggests that the Universe is described by quantum chaos 
during the dS era, with a corresponding scrambling time scale and Lyapunov exponent.
Interestingly, the complexity during this era appears to be unbounded, and will continue to grow linearly with the number of e-folds for as long as dS expansion continues.
When the dS expansion is followed by a period of radiation domination the complexity decreases
until ``freezing in'' once the mode re-enters the horizon.

We believe this new approach will open up the possibility of many future research directions. 
One obvious extension is to apply our analysis for other cosmological scenarios and models; for example, it would be interesting to study the complexity for accelerating solutions different from dS, or the complexity for hydrodynamical perturbations with sound speeds different than one. 
We also found that the complexity for a mode that exits the horizon during dS then re-enters the horizon during radiation initially increases, then decreases and ``freezes-in'' after horizon re-entry.
Since complexity represents the number of unitary quantum gates necessary to build the target state from the reference state, this suggests that there may be some sort of ``short cut'' in the space of quantum operators that can encode the spectrum of cosmological perturbations upon horizon re-entry.
As another potential application, we found that the complexity during the dS era grows linearly with the number of e-folds without bound, at a rate proportional to the dS Hubble expansion.
However, it has been suggested that the complexity for a system with a fixed number of q-bits should be bounded from above, and that the rate of growth of complexity should be bounded as well.
While these expectations appear to apply primarily to systems with time-independent Hamiltonians, it would be interesting to find connections between these ideas and cosmology, potentially placing limits on either the number of e-folds of dS expansion or the dS Hubble rate from quantum information theoretic grounds.

Finally, these results may be useful for understanding complexity in simple quantum optics setups, where the squeezed vacuum state arises quite naturally.
We would like to explore these potential directions in the near future. 

\section*{Acknowledgements}

We would like to thank Jeff Murugan for reading the manuscript and comments. AB is supported by Research Initiation Grant (RIG/0300) provided by IIT-Gandhinagar. This work was supported by the Natural Sciences and Engineering Research Council of Canada. AB thank the organizers and the participants of the workshop hosted by  Department of Physics of Ashoka University, Sonipat, Haryana, India, on holography, complexity and entanglement and  National Strings Meeting 2019, hosted by the Department of Physics, IISER Bhopal, India, to give him the opportunity to present a talk on the complexity and for stimulating discussions.  

\appendix
\setcounter{section}{0}
\section{Circuit Complexity using Wavefunction}
\label{app:Complexity}

We briefly review circuit complexity. First we will directly use the wavefunction and compute the circuit complexity using Nielsen's method \cite{NL1,NL2,NL3}. 
In this section we will only provide a brief outline. The details can be found in \cite{MyersCC}. 

\par
The problem is the following: given a set of elementary gates and a reference state, what is the most efficient quantum circuit that starts at that reference state (at $s=0$)
and terminates at a target state ($s=1$).
\begin{equation}
    |\Psi_{s=1}\rangle = U (s=1) |\Psi_{s=0}\rangle,
\end{equation}
where $U$ is the unitary operator that takes the reference state to the target state. We construct it from a continuous sequence of parametrized path ordered exponential of the Hamiltonian operator 
\begin{equation}
U(s)= {\overleftarrow{\mathcal{P}}} e^{- i \int_0^{s} ds H(s)}.
\end{equation}
Here $s$ parametrizes a path in the space of the unitaries and given a set of elementary gates $M_I$, the Hamiltonian can be written as
\begin{equation}
    H(s)= Y(s)^{I} M_{I}\, .
\end{equation}
The coefficients $Y^I$ are the control functions that dictates which gate will act at a given value of the parameter. The control function is basically a tangent vector in the space of unitaries and satisfy the Schrodinger equation
\begin{equation}
\frac{d U(s)}{ds} = -i\, Y(s)^I M_I U(s)\,.
\end{equation}
Then we define a cost functional $\mathcal{F (U, \dot U)}$ as follows:
\begin{equation}
{\mathcal C}(U)= \int_0^1 \mathcal{F} (U, \dot U) ds\, .
\end{equation}
Minimizing this cost functional gives us the optimal circuit. There are different choices for the cost functional \cite{NL1,MyersCC1}; in this paper we will consider {\it linear} and {\it quadratic} cost functionals
\be
\label{linearCost}
\mathcal{F}_1(U, Y) &=& \sum_I \left|Y^I\right|\, ; \\
\mathcal{F}_2 (U, Y) &=& \sqrt{\sum_I (Y^I)^2}\, .
\label{quadCost}
\ee

In order to compute the complexity we need to clearly specify the target and reference states.
In the context of cosmological pertubations, we will choose our target state to be the two-mode squeezed vacuum state
\begin{equation}
    |\Psi_{sq}\rangle_{\vec{k}} = \hat {\mathcal S}(r_k,\phi_k)_{\vec{k}} |0\rangle_{\vec{k}} = \frac{1}{\cosh r_k} \sum_{n=0}^{\infty} (-1)^n e^{-2 i n \phi_k} \tanh^n r_k\, |n_{\vec{k}}; n_{-\vec{k}}\rangle\, .
    \label{psi1Appendix}
\end{equation}
The wavefunction for this state can be written as a gaussian \cite{book,Martin2}
\begin{equation}
\Psi_{sq}= \mathcal{N} e^{A(q_{\vec{k}}^2+q_{-\vec{k}}^2)-B q_{\vec{k}} q_{-\vec{k}}}\, .
\label{squeezeGaussianAppendix1}
\end{equation}
By a suitable rotation it is possible to diagonalize the exponent,
\begin{equation}
\Psi_{sq}=\mathcal{N} e^{-\frac{1}{2} \tilde M^{ab}q_a q_b},
\label{squeezeGaussianAppendix2}
\end{equation}
where 
\begin{equation}
\tilde M=
\begin{pmatrix}
-2 A+B & 0 \\
0 & -2A-B
\end{pmatrix} \equiv \begin{pmatrix}
\Omega_{\vec{k}} & 0 \\
0 & \Omega_{-\vec{k}}
\end{pmatrix}.
\label{gaussianMatrix}
\end{equation}

Our reference state is the unsqueezed vacuum, which also has a gaussian wavefunction
\begin{equation}
    \psi_{R}= {\mathcal N} e^{-\frac{k}{2} (q_{\vec{k}}^2+ q_{-\vec{k}}^2)}={\mathcal N} e^{-\frac{1}{2} \sum_{k,-k} \omega_{\vec{k}} q_{\vec{k}}^2},
    \label{referenceGaussianAppendix}
\end{equation}
where $\omega_{\vec{k}}=k$. 
Using (\ref{squeezeGaussianAppendix2}) and (\ref{referenceGaussianAppendix}), we can compute expressions for the complexity using the two different cost functions 
(\ref{linearCost}),(\ref{quadCost}) \cite{NL1,MyersCC,MyersCC1}
\begin{eqnarray}
\label{complex1}
    {\mathcal C}_1(k) &=&\frac{1}{2}  \left (\ln \left|\frac{\Omega_{\vec{k}}}{\omega_{\vec{k}}}\right|+  \ln \left|\frac{\Omega_{-\vec{k}}}{\omega_{-\vec{k}}}\right|+ \tan^{-1} \frac{\text{Im}\ \Omega_{\vec{k}}}{{\text{Re}\ \Omega_{\vec{k}}}} + \tan^{-1} \frac{\text{Im}\ \Omega_{-\vec{k}}}{{\text{Re}\ \Omega_{-\vec{k}}}} \right)\, ; \\ \cr
    {\mathcal C}_2(k) &=&\frac{1}{2} \sqrt{ \left(\ln \left|\frac{\Omega_{\vec{k}}}{\omega_{\vec{k}}}\right| \right)^2+ \left(\ln \left|\frac{\Omega_{-\vec{k}}}{\omega_{-\vec{k}}}\right| \right)^2+ \left(\tan^{-1} \frac{\text{Im}\ \Omega_{\vec{k}}}{\text{Re}\ \Omega_{\vec{k}}}\right)^2+ \left(\tan^{-1} \frac{\text{Im}\ \Omega_{-{\vec{k}}}}{\text{Re}\ \Omega_{-{\vec{k}}}}\right)^2},
\label{complex2}
\end{eqnarray}
where $\Omega_k,\Omega_{-k}$ are defined as in (\ref{gaussianMatrix}). The inverse tangent term in the above expression is necessary when the frequency is complex \cite{AB}. This is the primary result of this section; the expressions (\ref{complex1}),(\ref{complex2}) will be used in Section \ref{sec:Complexity} to compute the complexity for the two-mode squeezed vacuum for cosmological perturbations.

\section{Circuit Complexity using Covariance matrix}
\label{app:Covariance}

The two-mode squeezed vacuum state (\ref{state1}) in Section \ref{sec:Complexity} is a Gaussian state and can  equivalently be described by a covariance matrix.
The covariance matrix takes the following form,  
\be
G^{s=1}_k=\left(
\begin{array}{cccc}
 \frac{1}{ \text {Re}  (\Omega_{\vec{k}})}& -\frac{\text {Im}  (\Omega_{\vec{k}})}{\text {Re}  (\Omega_{\vec{k}})}&0&0 \\
-\frac{ \text {Im}  (\Omega_{\vec{k}})}{\text {Re}  (\Omega_{\vec{k}})} & \frac{|\Omega_{\vec{k}} |^2}{\text {Re}  (\Omega_{\vec{k}})}&0& 0\\
0&0&\frac{1}{ \text {Re}  (\Omega_{-\vec{k}})}& -\frac{\text {Im}  (\Omega_{-\vec{k}})}{\text {Re}  (\Omega_{-\vec{k}})}\\0&0&-\frac{ \text {Im}  (\Omega_{-\vec{k}})}{\text {Re}  (\Omega_{-\vec{k}})} & \frac{|\Omega_{-\vec{k}} |^2}{\text {Re}  (\Omega_{-\vec{k}})}\end{array}
\right),
\ee
where $\Omega_{\vec{k}}=-2 A+B, \Omega_{-\vec{k}}=-2A-B$ are defined below (\ref{complexity2}). For the reference state this matrix will take the following form,
\be
G^{s=0}_k=\left(
\begin{array}{cccc}
 \frac{1}{k}& 0 &0&0\\
0 & k&0&0 \\
0&0&\frac{1}{k}&0\\
0&0&0&k\\
\end{array}
\right).
\ee

The complexity can then easily be computed to be \cite{AB}
\begin{align}
\begin{split} \label{answ}
\mathcal{C}_2(k)=\frac{1}{2}\sqrt{\left(\cosh^{-1}\left[\frac{k^2+|\Omega_{\vec{k}}|^2}{2\,k\text {Re}  (\Omega_{\vec{k}})}\right]\right)^2+\left(\cosh^{-1}\left[\frac{k^2+|\Omega_{-\vec{k}}|^2}{2\,k\text {Re}  (\Omega_{-\vec{k}})}\right]\right)^2} .
\end{split}
\end{align}
Using the explicit forms for $\Omega_{\vec{k}},\Omega_{-\vec{k}}$ from (\ref{complexity2}), this simplifies to be
\be
{\mathcal C}_2(k) = 2\sqrt{2}\ r_k(t)\, .
\ee
Interestingly, this result for the complexity from the covariance matrix method is independent of the squeezing angle $\phi_k$, while the circuit complexity computed for both the inverted harmonic oscillator in Section \ref{sec:Inverted} and cosmological perturbations in Section \ref{sec:Complexity} both depend on the squeezing angle.

In particular, the covariance matrix complexity would indicate that the complexity of the cosmological perturbations relative to the ground state from Section \ref{sec:Complexity} would continue to grow (as the number of e-folds) as long as the squeezing parameter $r_k$ grows, even in the radiation era where we found that complexity decreased for the circuit complexity. As already noted in \cite{AB}, however, the covariance matrix method of computing complexity appears to be less sensitive to fine details of the wavefunction (such as the squeezing angle), and the circuit complexity is a more precise measure of the complexity using wavefunction of a target state.

      

\end{document}